\documentclass{ws-procs9x6}
\def\GeV{{\rm GeV}}

\begin{document}

\title{The QCD Spin Structure of Nucleons\\
(Summary of Parallel Session 2)}
%Proceedings\footnote{\uppercase{T}his work is supported by etc, etc.}}

\author{G.~K. MALLOT}
%\footnote{\uppercase{W}ork partially supported by grant 2-4570.5 of the \uppercase{S}wiss 
%\uppercase{N}ational \uppercase{S}cience \uppercase{F}oundation.}}

\address{CERN,\\
1211 Geneva 23, Switzerland,\\
%World Scientific Publishing Co., Inc, \\
%1060 Main Street, \\ 
%River Edge, NJ 07661, USA\\ 
E-mail: gerhard.mallot@cern.ch}

\maketitle

\abstracts{
This paper attempts to summarise the highlights of the talks presented
in Parallel Session~II of the SPIN~2004 Symposium dedicated to the QCD
spin structure of nucleons. Emphasis is put on new data and theoretical
developments.
}

\section{Introduction}
The enormous interest in the spin structure of nucleons is 
reflected in the 55 accepted parallel talks in this session, comprising
about an even number of theoretical and experimental presentations.
About two-thirds of the experimental results came
from lepton--nucleon scattering, while the rest were from pp interactions.
A focus of the theory contributions was on transversity and single-spin asymmetries.
%Of particular interest were the first results from COMPASS at CERN and from the 
%RHIC-Spin Collaboration at BNL.
% as well as the continued flow of data from HERMES.
This paper will not attempt to cover the excellent plenary talks on spin 
structure nor can it do justice to all the excellent talks in the parallel session.

\begin{figure}[ht]
\begin{minipage}[t]{0.48\hsize}
\hbox to 0pt{~}% needed for correct positioning of minipage!
\vbox to 8cm{
\centerline{\epsfxsize=\hsize\epsfbox[13 0 509 325]{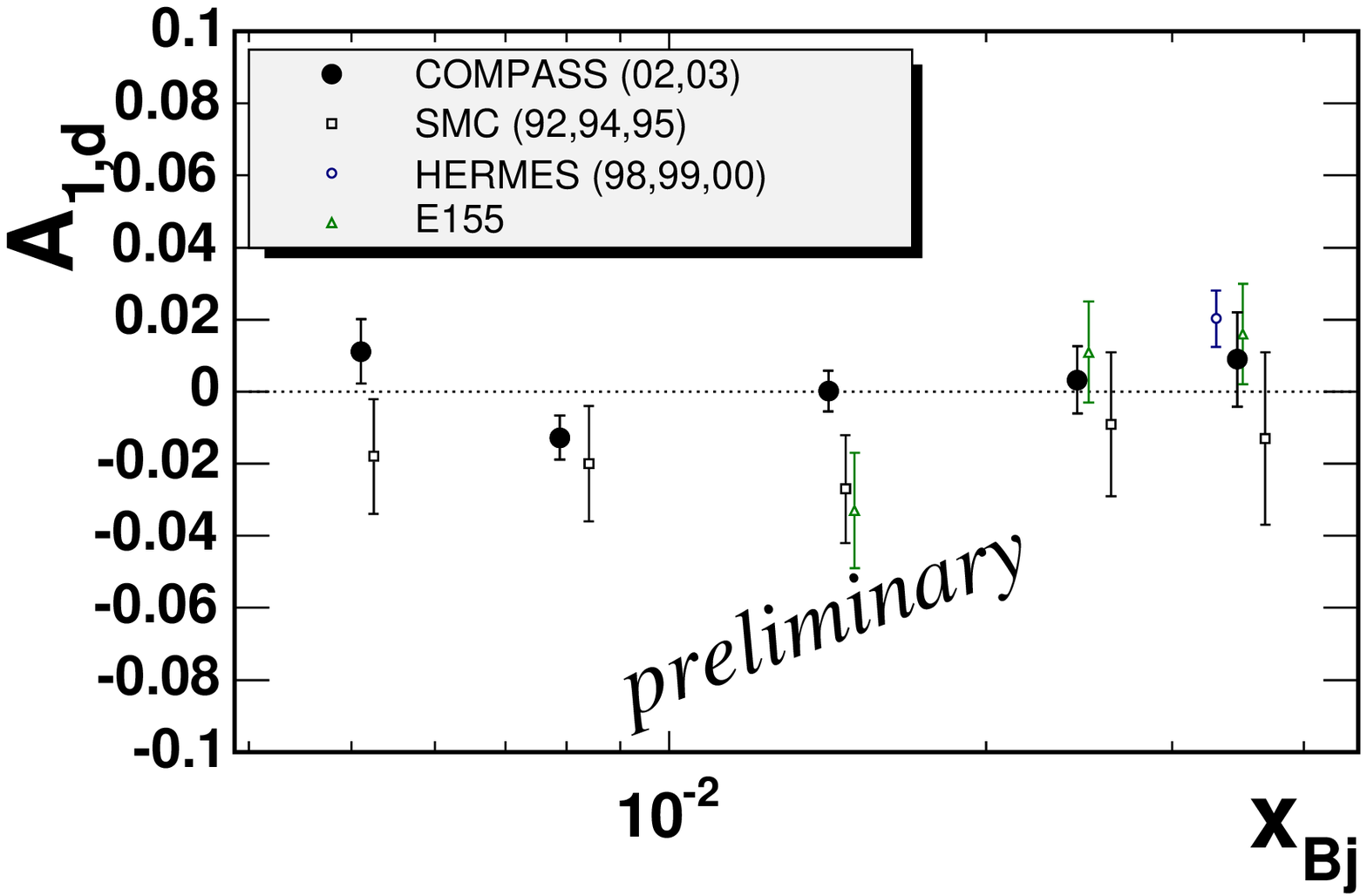}}   
%\caption{%\protect\raggedright
%$A_1^{\rm d}$ data in the small-$x$ region including the new COMPASS data.\protect\cite{Peshekhonov}
%\label{fig:C_a1d}}
%
%\addtocounter{figure}{1}
\vspace*{3mm}
\centerline{\epsfxsize=\hsize\epsfbox[8 0 510 325]{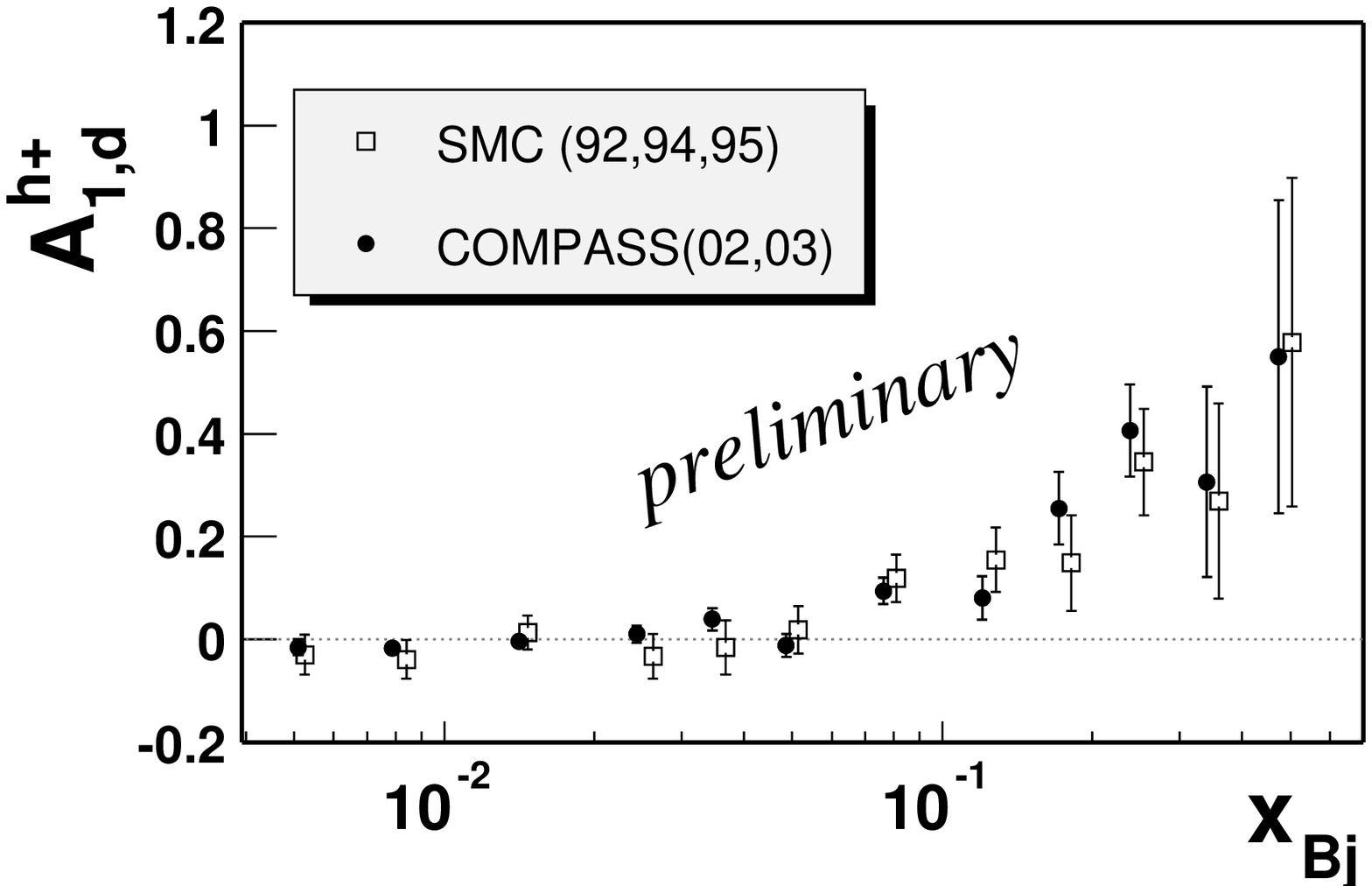}}   
}
\caption{%\protect\raggedright
Virtual photon asymmetry $A_1^{\rm d}$ for the deuteron from COMPASS 2002 and 2003 data:\protect\cite{Peshekhonov}
inclusive asymmetry for $x<0.04$ (top), asymmetry for positive hadrons (full $x$-range, bottom). 
\label{fig:C_a1d}}
\end{minipage}
\hfill
\begin{minipage}[t]{0.48\hsize}
\hbox to 0pt{~}
\vbox to 8cm{
%\addtocounter{figure}{-2}
%\vfil
\centerline{\epsfxsize=\hsize\epsfbox{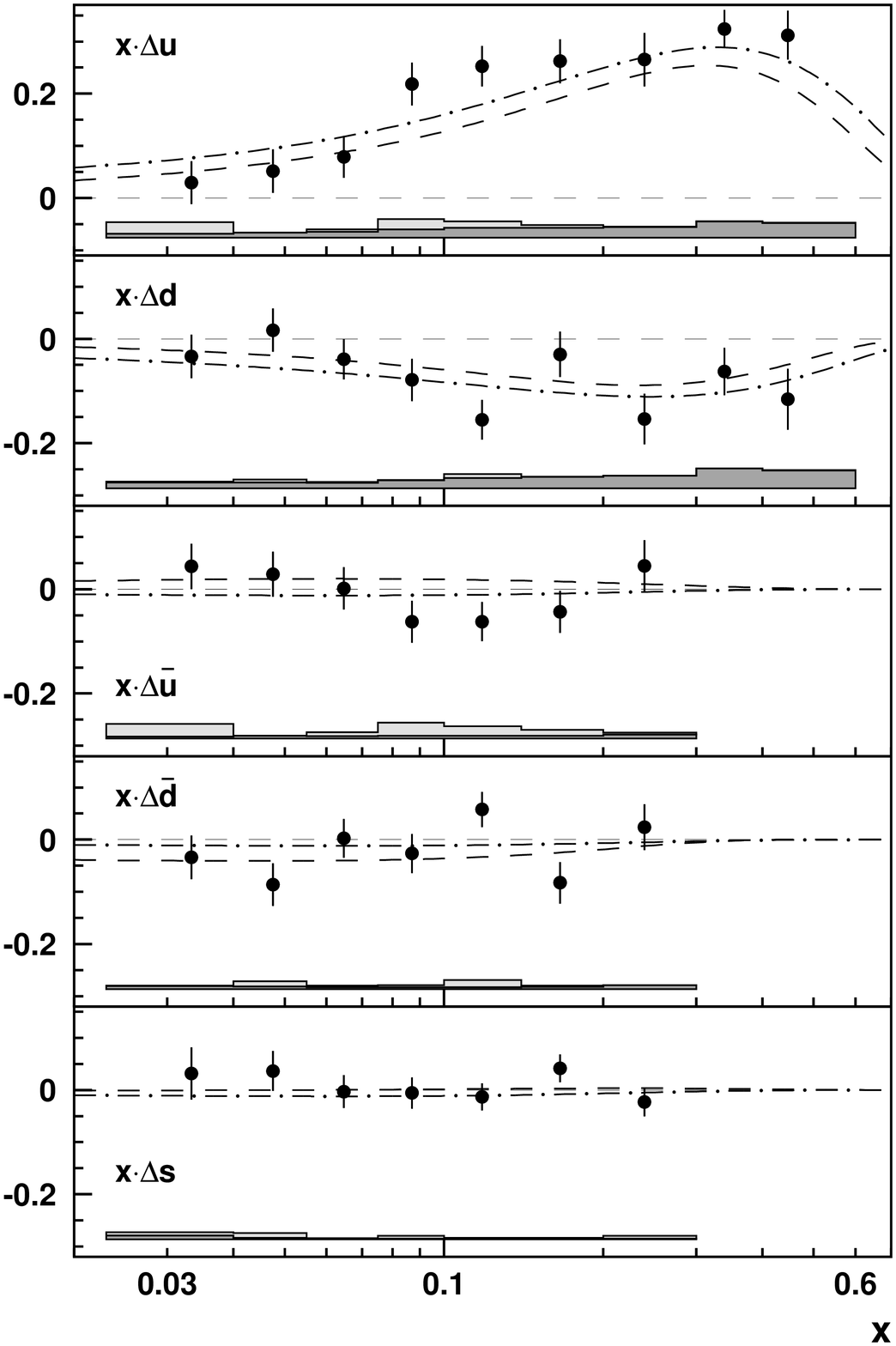}}
}
\caption{%\protect\raggedright
The quark helicity distributions $x\Delta q(x)$ at fixed $Q^2=2.5~\GeV^2$ from HERMES.\protect\cite{Rubin}
The strange quark distribution $x\Delta s$ was determined assuming $x\Delta\overline{s}\equiv0$.
\label{fig:H_dq}}
%\addtocounter{figure}{1}
\end{minipage}
\end{figure}

\section{Quark Helicity Distributions}
The spin-dependent structure functions $g_1(x,Q^2)$ of the proton, the deuteron and 
the neutron are well known by now. However, the lack of data from colliders at small 
$x$ and large $Q^2$, limit the information obtainable from an analysis of the $Q^2$
evolution. HERMES at DESY completed its programme with longitudinal
polarisation and presented their quasi-final results\cite{Riedl} for $g_1$ of the proton, the
deuteron and the neutron.
%, corrected for smearing of the kinematic variables. 
COMPASS showed its first $A_1$ deuteron data\cite{Peshekhonov} from the 2002 and 2003 runs.
For $x<0.03$ these are the most precise data yet (Fig.~\ref{fig:C_a1d}). The somewhat negative
tendency of the SMC data in this region is not reproduced.
During 2004 about the same amount of data was recorded and the larger-$x$ region will
additionally profit from an improved trigger set up.

In the valence-quark region $x>0.2$ new highly precise data became available from Jefferson Lab.
The $^3$He neutron data\cite{E99-117} from E99-117 show for the first time a clearly
positive asymmetry $A_1^{\rm n}$ for $x\simeq0.6$. The deviation from pQCD predictions
may hint to a possible effect due to quark orbital angular-momentum.\cite{E99-117}

\begin{figure}[ht]
\begin{minipage}[t]{0.48\hsize}
\hbox to 0pt{~}
%\vbox to 6cm{
%\vfill
\centerline{\epsfxsize=\hsize\epsfbox{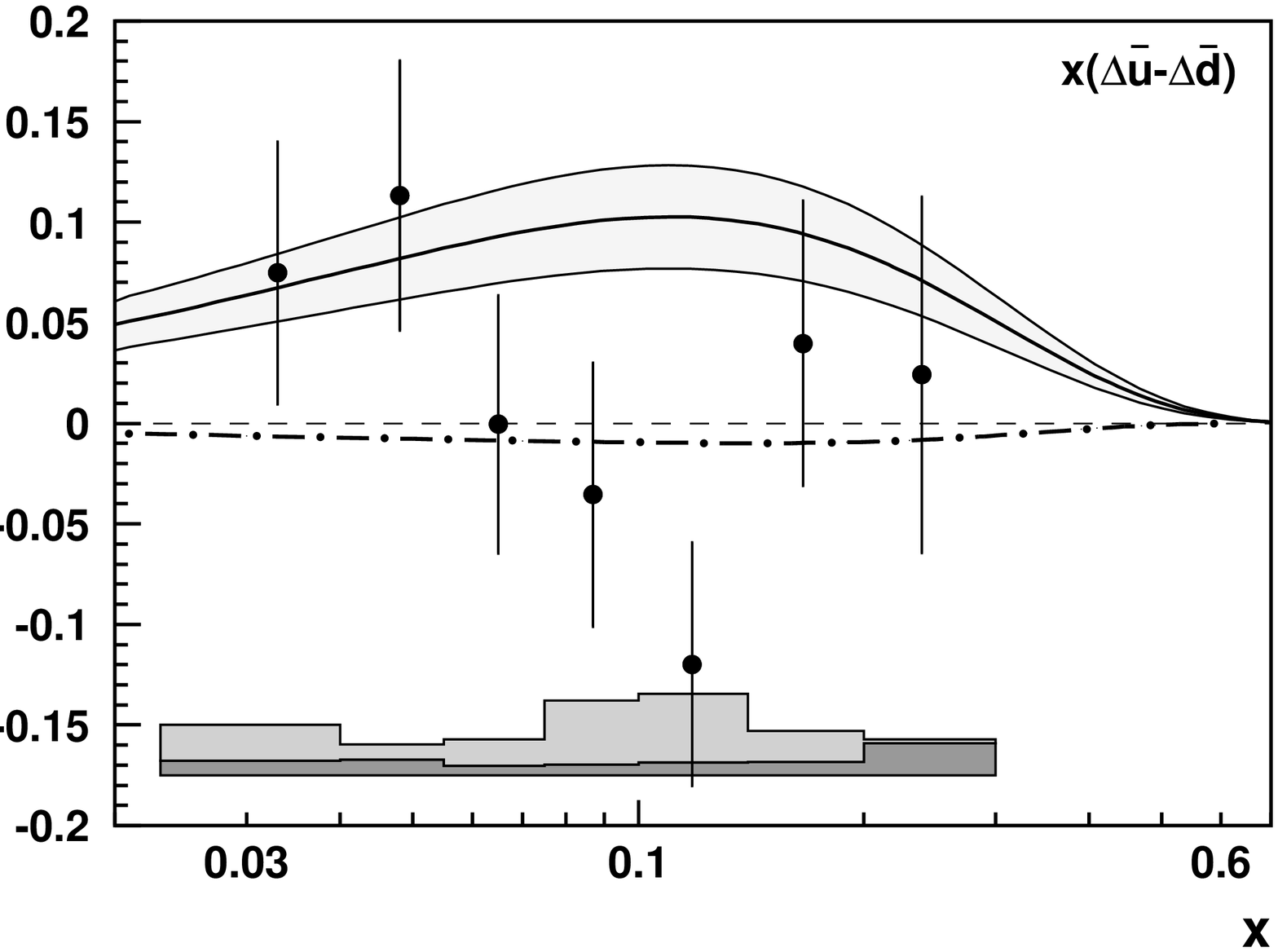}}
%}
\caption{%\protect\raggedright
The difference $x(\Delta \overline{u} -\Delta \overline{d})$ as function of $x$ from 
HERMES\protect\cite{Rubin} at $Q^2=2.5~\GeV^2$.
\label{fig:H_udbar}}
\end{minipage}
\hfill
\begin{minipage}[t]{0.48\hsize}
\hbox to 0pt{~}
\vbox to 6cm{
\centerline{\epsfxsize=\hsize\epsfbox{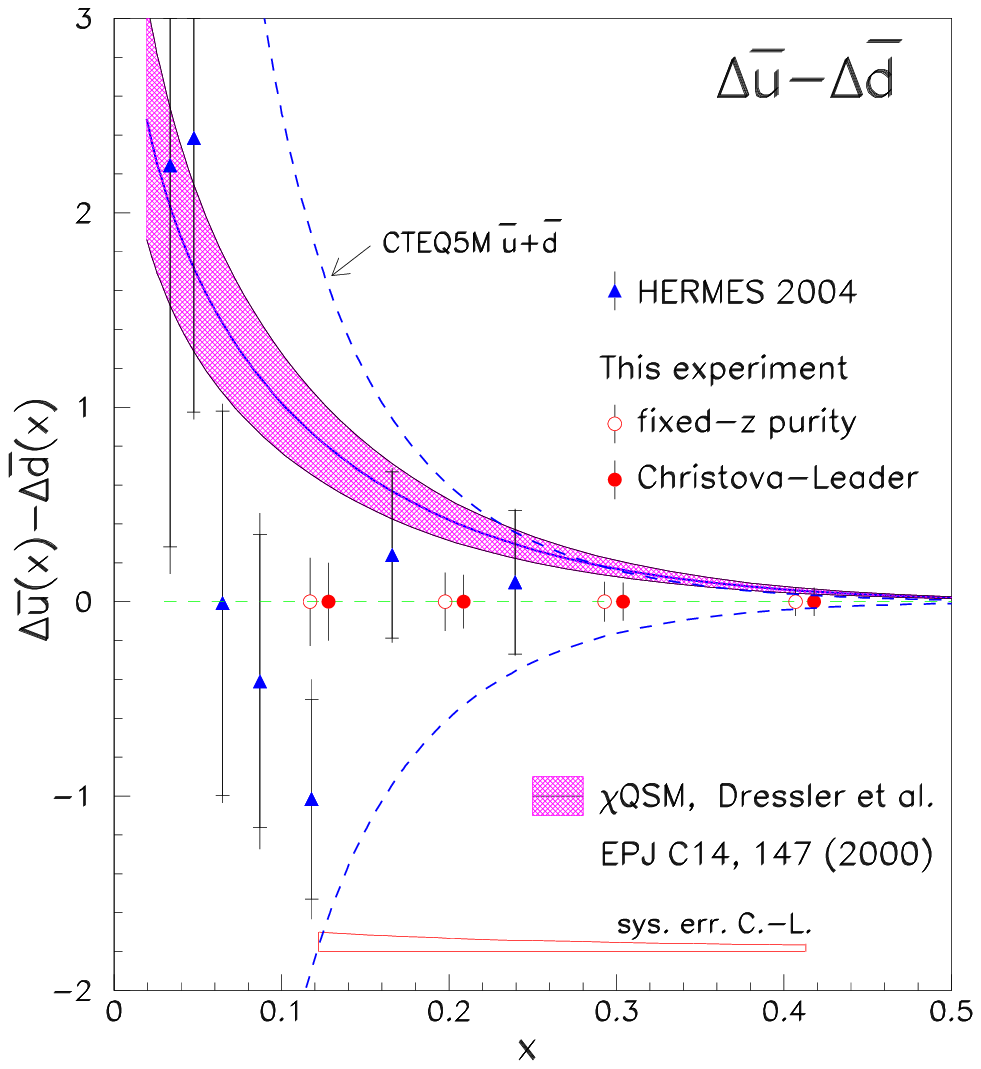}}
}
\caption{%
Results for $\Delta\overline{u}(x)-\Delta\overline{d}(x)$ 
%the future 
expected from Semi-SANE.\protect\cite{Jiang}
\label{fig:semi_sane}}
\end{minipage}
\end{figure}

To disentangle the contributions of the various quark flavours to $g_1$ requires semi-inclusive 
deep inelastic scattering, where the detected final-state hadron `remembers' the
flavour of the originally hit quark.  In particular the interesting polarisation of the 
strange quarks cannot be determined by inclusive DIS off proton and neutron (deuteron) 
targets. 
New semi-inclusive asymmetries were presented by COMPASS.\cite{Peshekhonov}
In Figure~\ref{fig:C_a1d} (bottom) the asymmetries for positive hadrons are shown reaching down
to $x\simeq 5\cdot10^{-3}$ for $Q^2>1~\GeV^2$. Similar results exist for negative hadrons, while 
the kaon asymmetries are still being analysed.
HERMES analysed its data, including the asymmetries from identified kaons, in terms of the 
five quark distributions:\cite{Rubin} 
$\Delta u$, $\Delta d$, $\Delta\overline{u}$, $\Delta\overline{d}$ and 
$\Delta s$ at fixed $Q^2=2.5~\GeV^2$ and assuming $\Delta\overline{s}\equiv0$.
The up and down quark data confirm, with higher precision, earlier results.
The up and down antiquark distributions as well as that of the strange quarks are compatible 
with zero (Fig.~\ref{fig:H_dq}).

It is well known that there is a strong isospin asymmetry\cite{udbar} in the unpolarised quark 
distributions $\overline{u}(x)-\overline{d}(x)\ne0$, 
first observed as a violation of the Gottfried sum rule. 
The models describing this asymmetry predict also an isospin asymmetry for
 $\Delta\overline{u}(x)-\Delta\overline{d}(x)$.
The HERMES data\cite{Rubin}
are shown in Fig.~\ref{fig:H_udbar} with two model predictions.\cite{udbar_mod} The data 
rather favour a symmetric sea.
High precision data for 
$\Delta\overline{u}(x)-\Delta\overline{d}(x)$ with $x>0.1$ are expected from a new JLAB experiment,
semi-SANE (E04-113).\cite{Jiang} In Figure~\ref{fig:semi_sane} the projected results are
shown. The scattered electron is detected in a calorimeter, while  the final-state hadron is 
detected in a magnetic spectrometer.
The inversion of the magnetic field yields the same acceptance
for oppositely charged hadrons while leaving the scattered-electron acceptance untouched.
This opens the way to using a model-independent NLO method\cite{Christova} involving positive and
negative hadron data, in which
%. In the ratios
%$$
%A^{h^+-h^-}_N=\frac{\Delta\sigma_N^{h^+}-\Delta\sigma_N^{h^-}}{\sigma_N^{h^+}-\sigma_N^{h^-}}
%\stackrel{N=\Pp, h=\pi}{=}\frac{4\Delta u_V-\Delta d_V}{4u_V-d_V}
%$$
the fragmentation functions cancel.
% and they become functions of only the valence-quark helicity
%distributions. 
A method using the first moments of the quark helicity distributions 
to determine $\Delta\overline{u}(x)-\Delta\overline{d}$ was also discussed.\cite{Shevchenko}

\begin{figure}
\begin{minipage}[t]{0.48\hsize}
\hbox to 0pt{~}
  \centerline{\epsfxsize=\hsize\epsfbox[9 12 538 511]{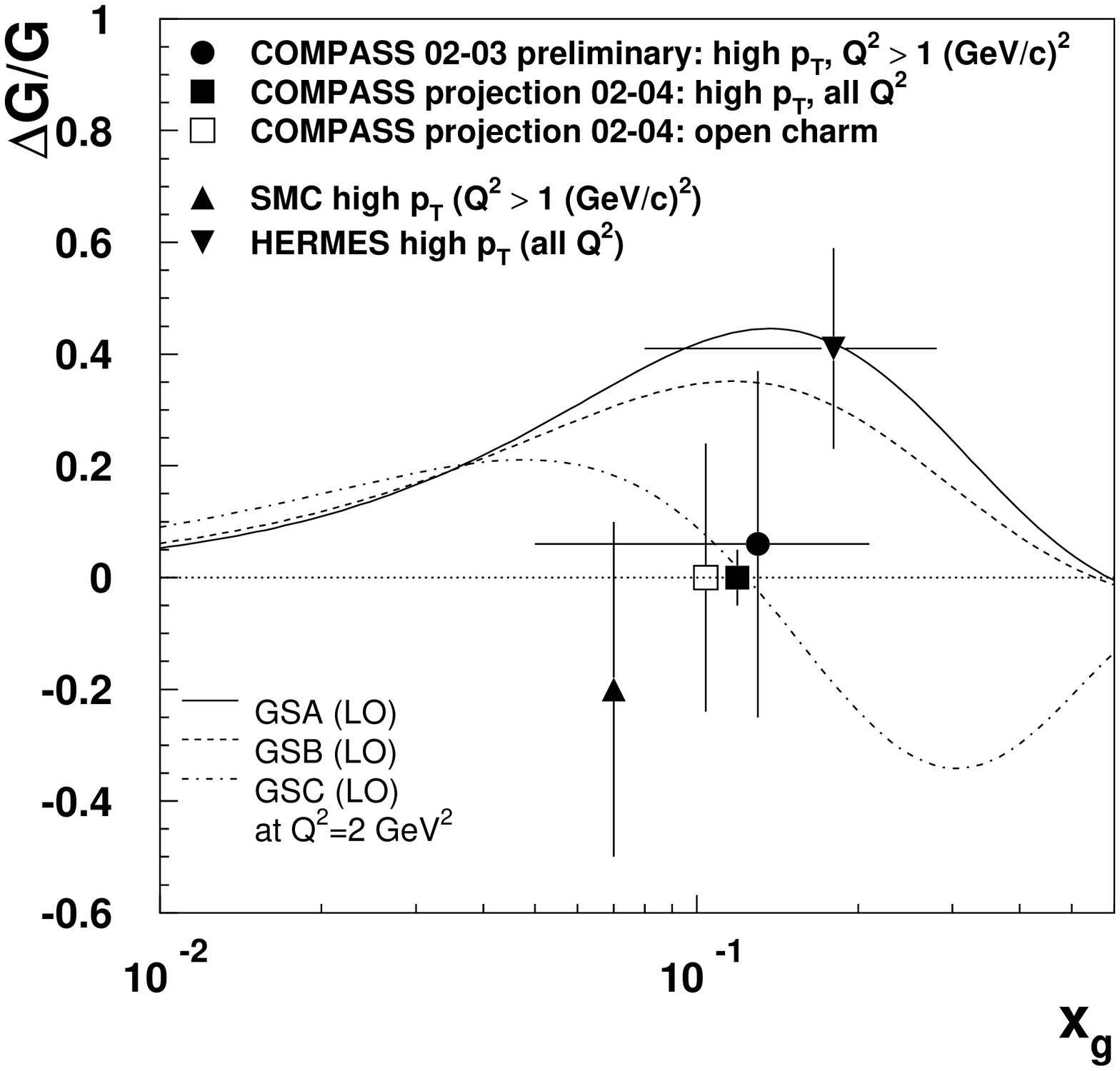}}
  \caption{%\protect\raggedright
    Gluon polarisation $\Delta G/G$ as function of $x$-gluon. The triangles and the circle correspond to 
    data\protect\cite{Rondio,Schill,H_pt} while the squares 
    indicate COMPASS projections.\protect\cite{Schill}
    % for the 2002--2004 data set based on present results. 
    %From Ref.~\protect\refcite{Schill}.
    \label{fig:delta_g}}
\end{minipage}
\hfill
\begin{minipage}[t]{0.46\hsize}
\hbox to 0pt{~}
\centerline{\epsfxsize=\hsize\epsfbox{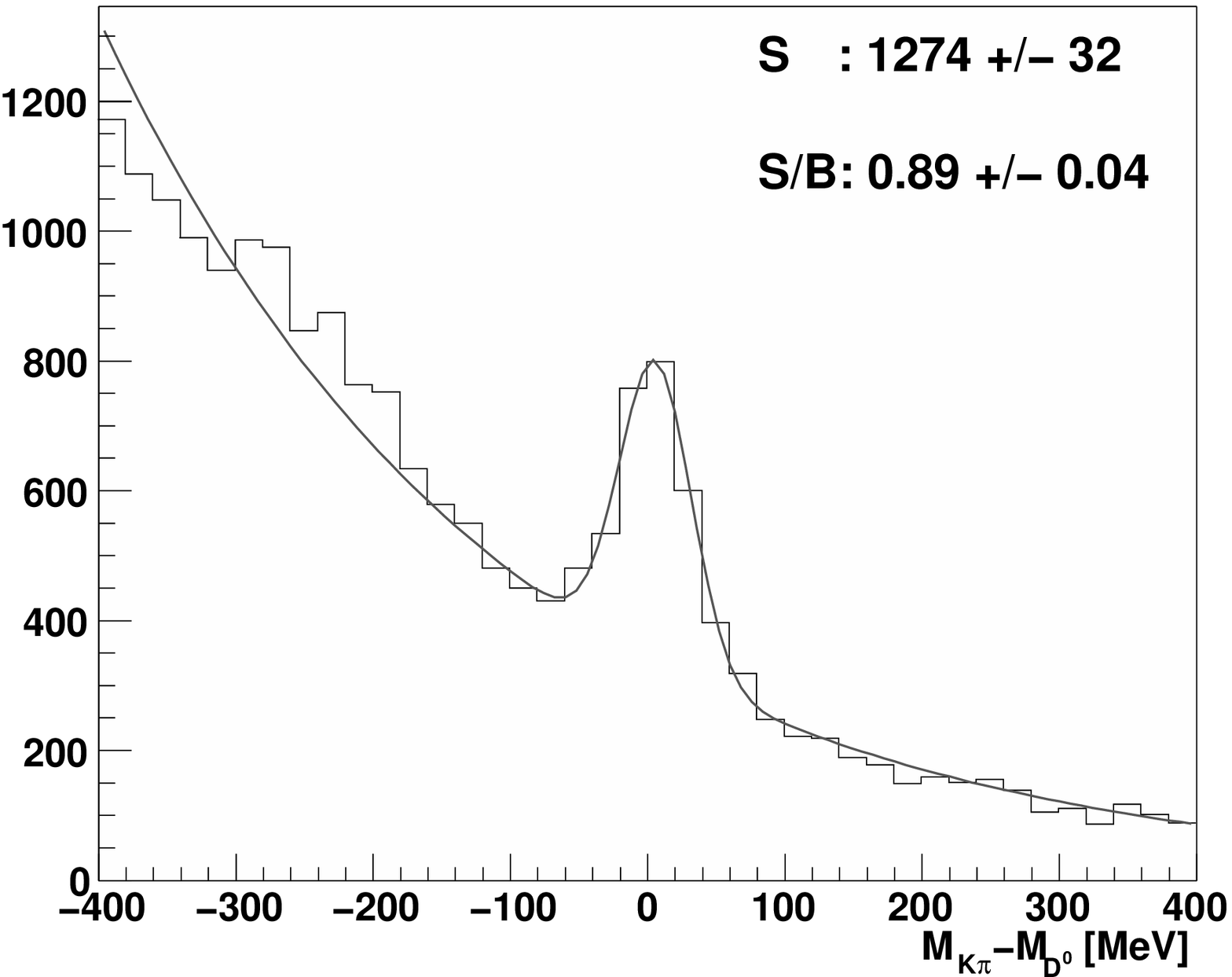}}
\caption{%\protect\raggedright
COMPASS invariant mass peak $M_{\rm K\pi}-m_{\rm D^0}$ from the
$\rm D^0 \rightarrow \pi K$ decay.\protect\cite{Schill}
\label{fig:d_peak}}
\end{minipage}
\end{figure}

Higher-twist effects, double logarithm resummation, and matrix solutions for PDFs were presented
in separate contributions.\cite{evol}

\section{Gluon Polarisation}

Still little is known about the gluon helicity distribution, $\Delta G$, which can in DIS either be determined
from the $Q^2$ evolution of the quark helicity distributions or from processes involving photon--gluon fusion
(PGF).
%(Fig.~\ref{fig:pgf_diag}c). 
The longitudinal double-spin asymmetry
%$A^{\ell N\rightarrow h h X}$ 
%
\begin{eqnarray}
A^{\ell N\rightarrow \ell'h h X}&=&\frac{\Delta G}{G}       \langle\hat a_{LL}\rangle^{PGF}  R^{PGF}\nonumber\\
                                    &+&\frac{\Delta q}{q}\left\{\langle\hat a_{LL}\rangle^{LP}   R^{LP}
                                     +                          \langle\hat a_{LL}\rangle^{QCDC} R^{QCDC}\right\}\nonumber
\end{eqnarray}
contains
terms involving $\Delta G/G$ and those involving $\Delta q/q$. The latter is rather well known experimentally
and the size of the relative contributions $R$ must be estimated by Monte Carlo simulations. The partonic asymmetries
$\langle\hat a_{LL}\rangle$ are known from theory.
\begin{figure}
\begin{minipage}[t]{0.48\hsize}
\hbox to 0pt{~}
\centerline{\epsfysize=4.5cm\epsfbox{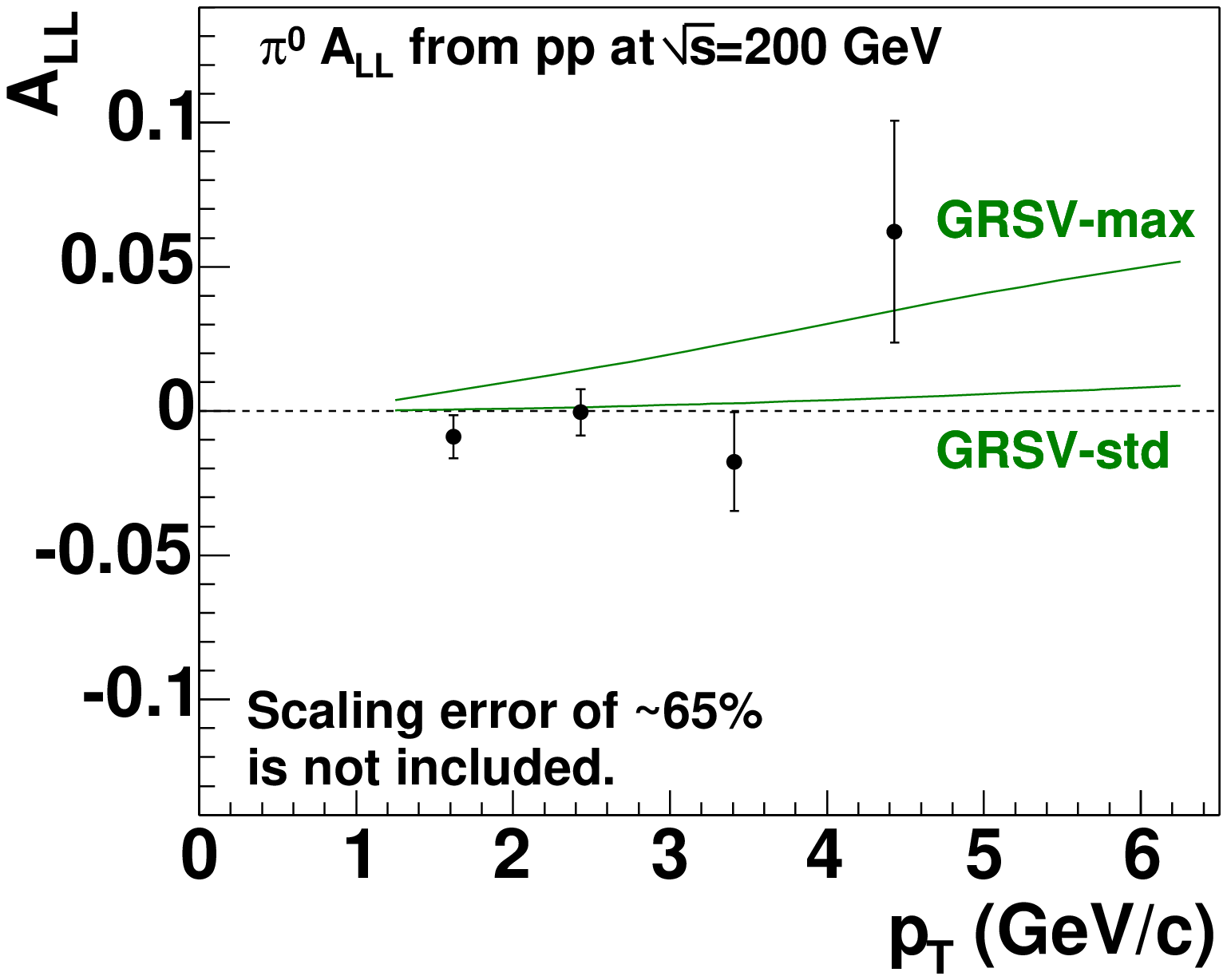}}
\caption{%\protect\raggedright
    Double-spin asymmetry $A_{LL}$ for $\pi^0$ production as a function of $p_T$ from 
    PHENIX.\protect\cite{Fukao}
%   2003 and 2004 data.\protect\cite{Fukao} 
    The curves correspond to the respective GRSV\protect\cite{grsv} input gluon-distributions.
    \label{fig:phenix_aLL_2004}}
\end{minipage}
\hfill
\begin{minipage}[t]{0.48\hsize}
\hbox to 0pt{~}
\centerline{\epsfysize=4.5cm\epsfbox{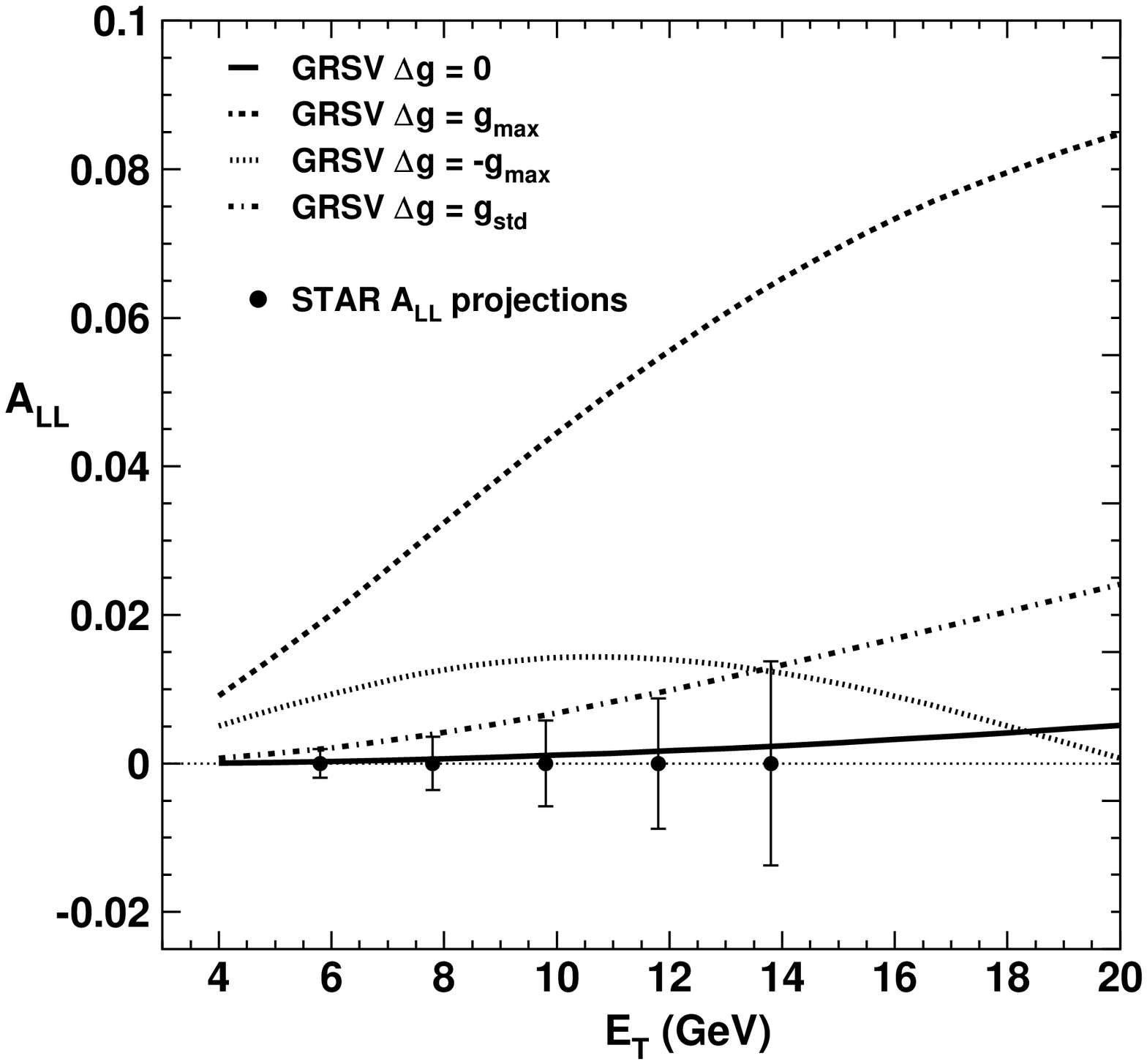}}
\caption{%\protect\raggedright
    Projected data for $A_{LL}$ from ${\rm pp\rightarrow jet}+X$ at STAR. The curves correspond
    to GRSV:\protect\cite{grsv}
    $g_{max}$, $g_{std}$, $g=0$, $-g_{max}$ (top to bottom on right side).
    \label{fig:star_jet}}
\end{minipage}
\end{figure}
Determinations of $\Delta G/G$ from PGF using unidentified high-$p_T$ hadron pairs and $Q^2>1~\GeV^2$ were reported by SMC\cite{Rondio}
and by COMPASS\cite{Schill} (2002--2003 data). The results are shown in Fig.~\ref{fig:delta_g} together with a previous result
from HERMES\cite{H_pt} (all $Q^2$). Both new results are compatible with zero and smaller 
than the HERMES value. Hadron pairs at $Q^2<1~\GeV$ are 10 times more abundant at COMPASS but their analysis is more
model dependent and must take into account contributions from resolved photon processes. The cleanest signature for a PGF
process is open-charm in the final state. COMPASS showed the first mass peak for D~mesons (Fig.~\ref{fig:d_peak}).
The projected precision for $\Delta G/G$ from D$^0$ asymmetries and from all-$Q^2$ hadron pairs for their 2002--2004 data are 
also shown in Fig.~\ref{fig:delta_g}.

\begin{figure}
\begin{minipage}[t]{0.48\hsize}
\hbox to 0pt{~}
\centerline{\epsfxsize=\hsize\epsfbox{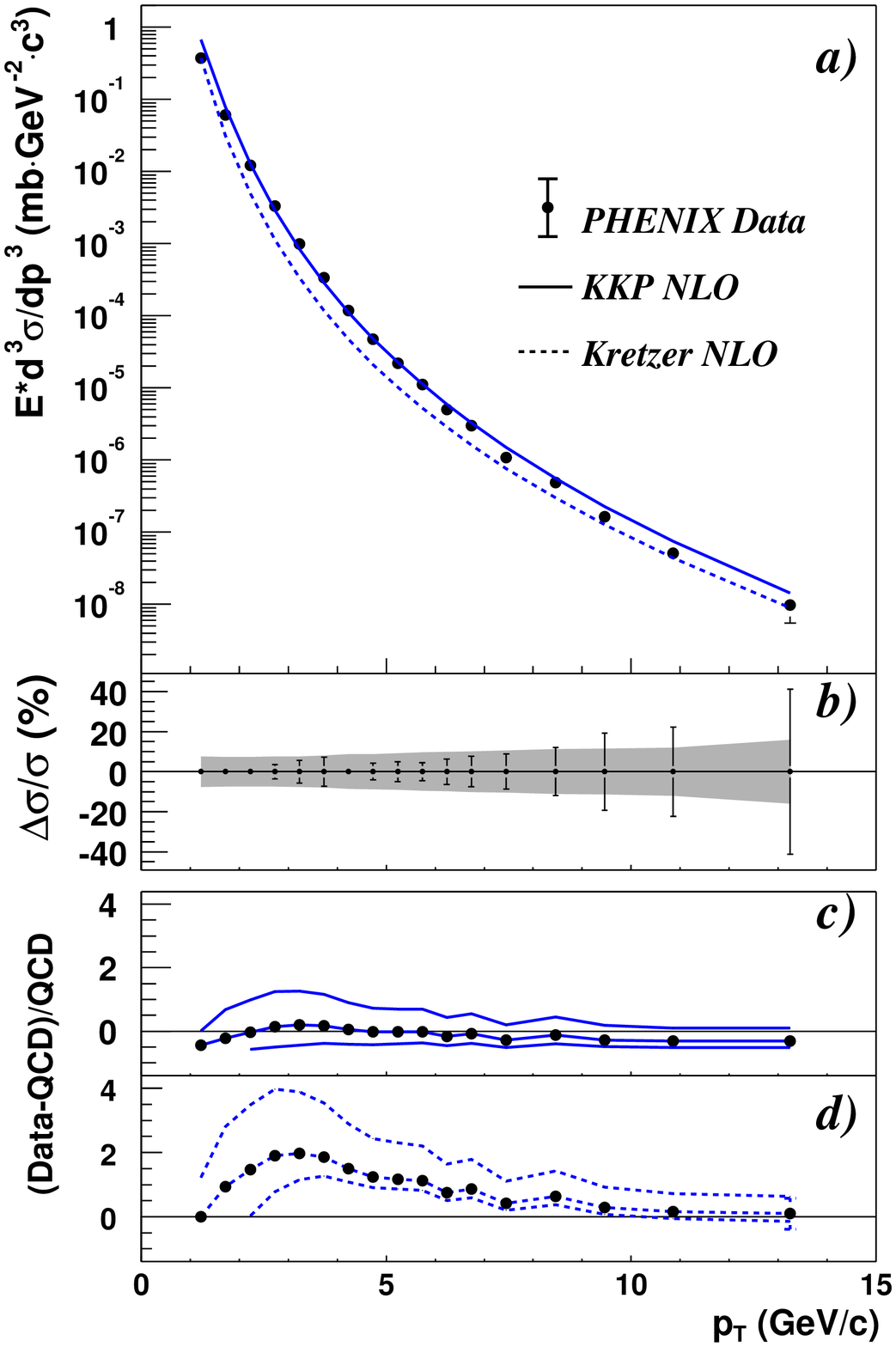}}
\caption{%\protect\raggedright
 The invariant differential cross-section for inclusive $\pi^0$ production from 
 PHENIX as function
 of $p_T$.\protect\cite{vogelsang,P_sig_pi0} For details see Ref.~\protect\refcite{P_sig_pi0}.
% from NLO pQCD calculations with equal renormalization and factorization scales of pT using the  
%Kniehl-Kramer-P\"otter (solid line) and  Kretzer (dashed line) sets of fragmentation functions. 
%b) The relative statistical (points) and point-to-point systematic (band) errors. c,d) The relative 
%difference between the data and the theory using KKP (c) and Kretzer (d) fragmentation functions 
%with scales of $p_T/2$ (lower curve), $p_T$ , and $2p_T$ (upper curve). In all figures, the normalization 
%error of 9.6\% is not shown. Figure taken from Ref.~\protect\refcite{P_sig_pi0}.
    \label{fig:phenix_sig_pi0}}
\end{minipage}
\hfill
\begin{minipage}[t]{0.48\hsize}
\hbox to 0pt{~}
\setbox1=\vbox{\epsfxsize=\hsize\epsfbox{phenix_sig_pi0.eps}}
\vbox to \ht1{
\centerline{\epsfxsize=\hsize\epsfbox{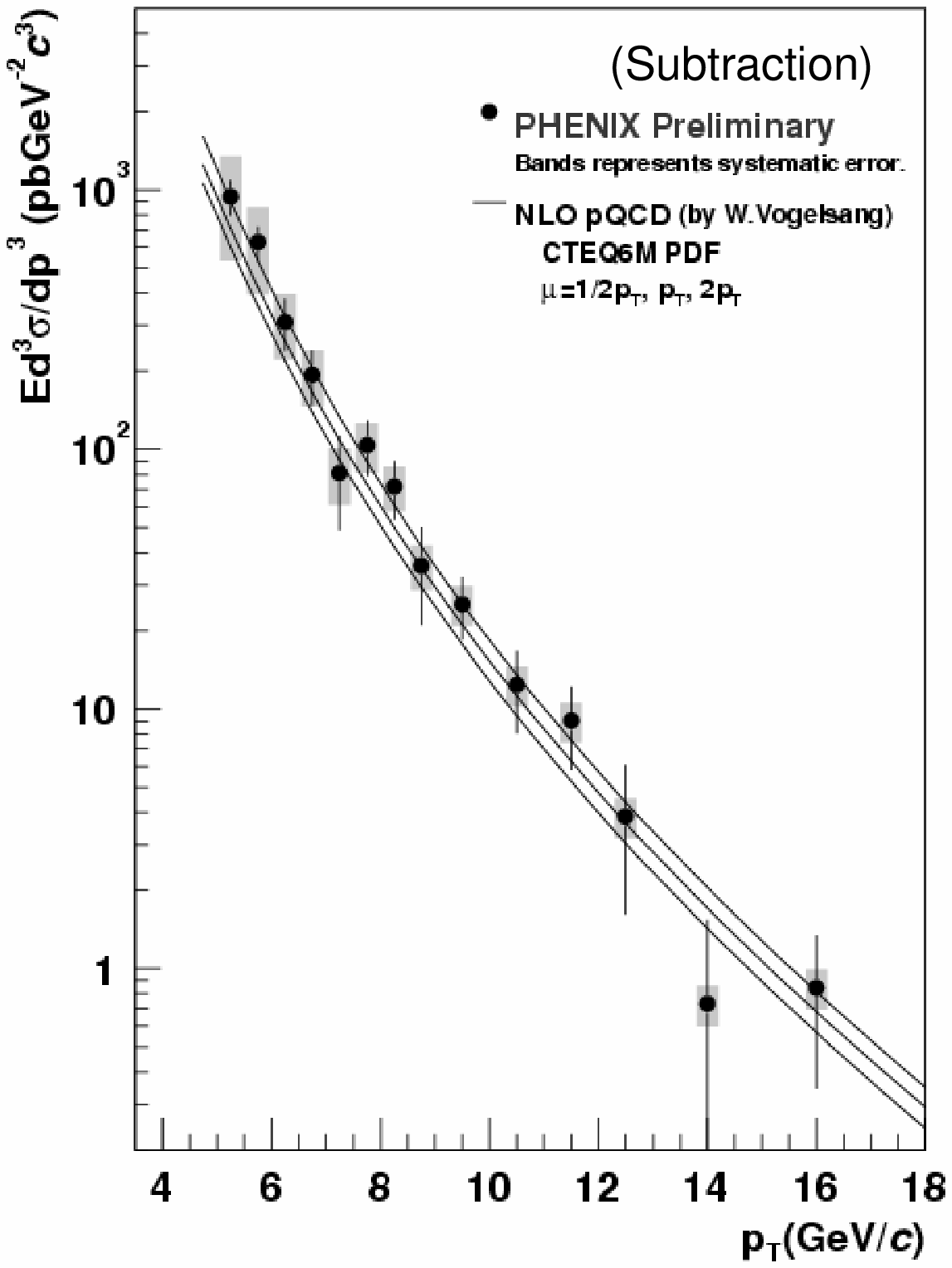}}
}
\caption{%\protect\raggedright
     The invariant differential cross-section for photon production from
     PHENIX as function of $p_T$.\protect\cite{okada}
     \label{fig:P_gamma}}
\end{minipage}
\end{figure}

The polarised proton--proton collisions at RHIC offer many channels to study
the gluon polarisation. 
Given the limited beam polarisation (2003) and beam time (2004) the most promising
channel is the double spin asymmetry in $\pi^0$ production, ${\rm pp\rightarrow\pi^0}X$.
The results\cite{Fukao} from PHENIX are shown in Fig.~\ref{fig:phenix_aLL_2004}.
The statistical significance is still very limited, but the big potential of this
measurement becomes apparent taking into account
the improvement in 2004 with respect to the 2003 data,\cite{P_pi0} 
which is largely due to the improved beam polarisation. 
% (Fig.~\ref{fig:phenix_aLL_2005}).
%The smallest asymmetry is generated for $\Delta G=0$ input at the input scale $\mu^2_{NLO}=0.40~\GeV^2$.
Also the STAR experiment expects significant data\cite{Sowinski} for the gluon polarisation from 
the process ${\rm pp}\rightarrow{\rm jet}+X$ from the 2005 run (Fig.~\ref{fig:star_jet}).
In pp collisions gluon--gluon, gluon--quark and
quark--quark partonic processes contribute yielding terms proportional to
$(\Delta G/G)^2$, $(\Delta G/G)(\Delta q/q)$ and $(\Delta q/q)^2$.  As a
consequence an ambiguity concerning the sign of $\Delta G$ arises.
Because of the quadratic term in $\Delta G$ there is almost no possibility to generate a
negative asymmetry\cite{jager} as illustrated in Figs.~\ref{fig:phenix_aLL_2004} and \ref{fig:star_jet}. 

Of central importance for the determination of $\Delta G/G$ from $\pi^0$ or prompt-photon production
at RHIC is that the cross-section is well understood and reproduced by theory. 
Figures~\ref{fig:phenix_sig_pi0} and \ref{fig:P_gamma} demonstrate the good level of agreement for 
collider c.m.s.\ energies. Next-to-leading logarithm resummation and power corrections are an important
element in the calculation.\cite{vogelsang,SteV_04a}

The determination of the first moment of $\Delta G$ requires an extension of the limited kinematic 
range in which the gluon momentum fraction $x_g$ is accessible with the present RHIC detectors. A considerable upgrade
programme\cite{P_upgrade} is proposed for PHENIX, which will extent the limits down to $x_g\simeq0.001$
at $\sqrt{s}=200~\GeV$.
Hyperon polarisation and new observables in pp collisions were also discussed.\cite{XuRykov}
%Q.~Xu discussed hyperon polarisation and V.L.~Rykov new observables in pp collisions.
\begin{figure}
\begin{minipage}[t]{0.48\hsize}
\hbox to 0pt{~}
\centerline{\epsfxsize=\hsize\epsfbox[78 50 570 536]{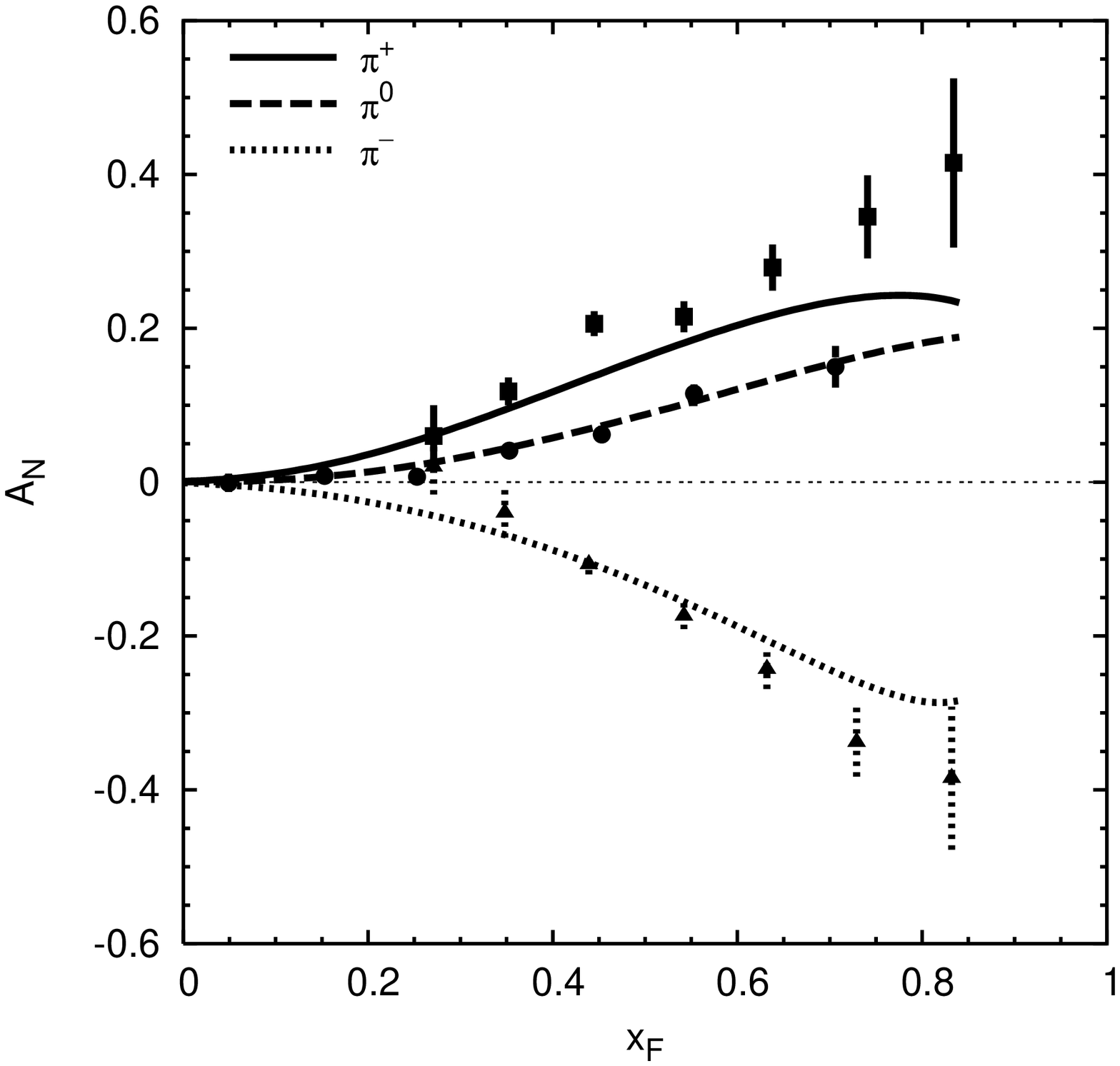}}
\caption{%
 E704 data\protect\cite{E704} described by the Sivers effect in the $k_T$ model.\protect\cite{murgia,AlM}
    \label{fig:E704_sivers}}
\end{minipage}
\hfill
\begin{minipage}[t]{0.48\hsize}
\hbox to 0pt{~}
\setbox1=\vbox{\epsfxsize=\hsize\epsfbox[78 50 570 536]{murgia_ssa.eps}}
\vbox to \ht1{
\centering{
\epsfxsize=\hsize\epsfbox{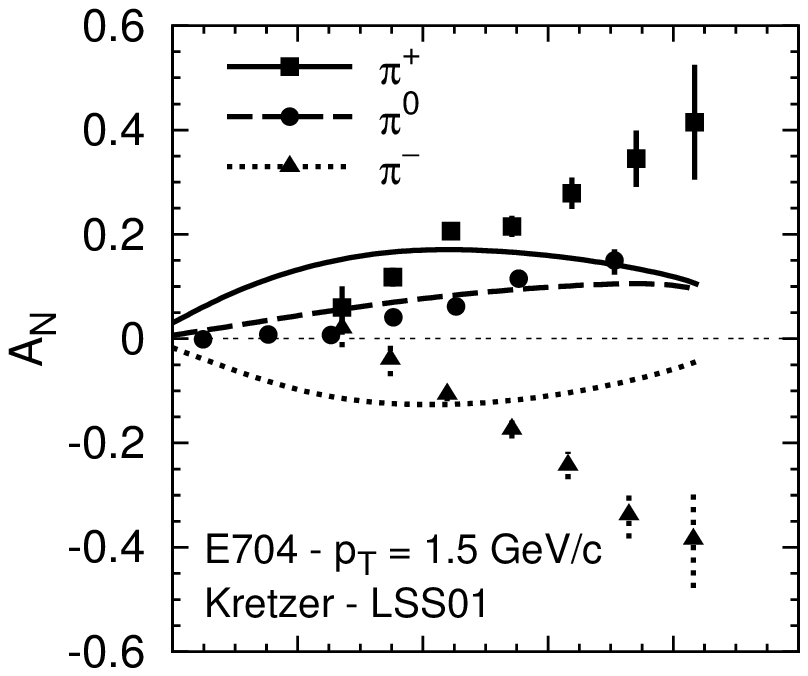}
\hbox{\hskip 1mm\epsfxsize=\hsize\epsfbox{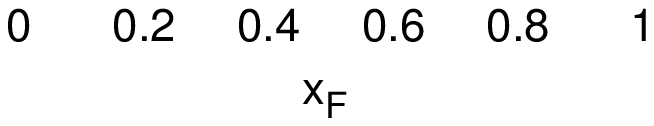}}
}
}
\caption{%
     Attempt to describe the E704 data\protect\cite{E704} purely by the Collins effect.\protect\cite{AlM,alesio}
    \label{fig:E704_collins}}
\end{minipage}
\end{figure}

\section{Transverse Spin}

New measurements of the single-spin transverse asymmetry $A_N$ for neutral pion production
were reported by PHENIX\cite{makdisi} for $x_F=0$ and by STAR\cite{ogawa} for 
$0.2<|x_F|<0.6$. The asymmetry is consistent with zero for $x_F<0.4$ and positive for
larger $x_F$. This behaviour is reproduced\cite{murgia} by the Sivers effect in a 
generalised leading-order pQCD model\cite{AlM} using $k_T$-dependent PDFs and
fragmentation functions. The model also describes the FNAL-E704 pion data\cite{E704} by
the Sivers effect (Fig.~\ref{fig:E704_sivers}). On the other hand even a fully saturated Collins 
mechanism\cite{alesio} cannot reproduce these data (Fig.~\ref{fig:E704_collins}). In the same model it was shown that D-meson production 
at RHIC in ${\rm p\!\uparrow p\rightarrow D}X$ is an ideal place to study the gluon 
Sivers-distribution.\cite{boglione} 
%Also Cahn asymmetries from EMC and Sivers asymmetries
Also the Sivers asymmetries
from HERMES\cite{schnell} can be described in this model.\cite{prokudin}

A full next-to-leading order calculation\cite{mukherjee} is now available for prompt-photon production
and other cross-sections in $\rm p\!\uparrow p\!\uparrow$ collisions. 
The scale dependence is strongly reduced as compared with LO calculations.

\begin{figure}
\begin{minipage}[t]{0.48\hsize}
\setbox1=\vbox{%
\centering{
\epsfxsize=\hsize\epsfbox{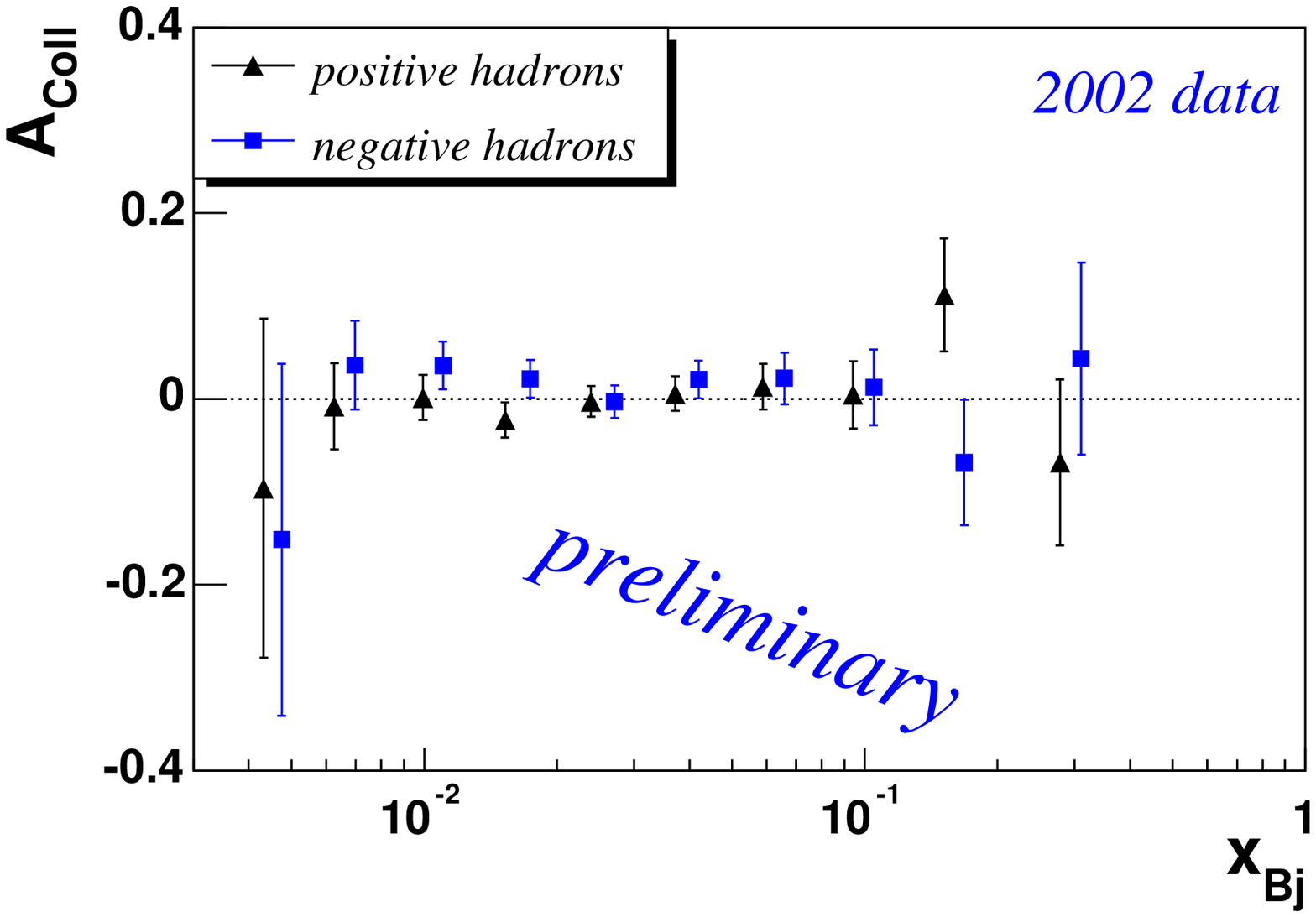}
\epsfxsize=\hsize\epsfbox{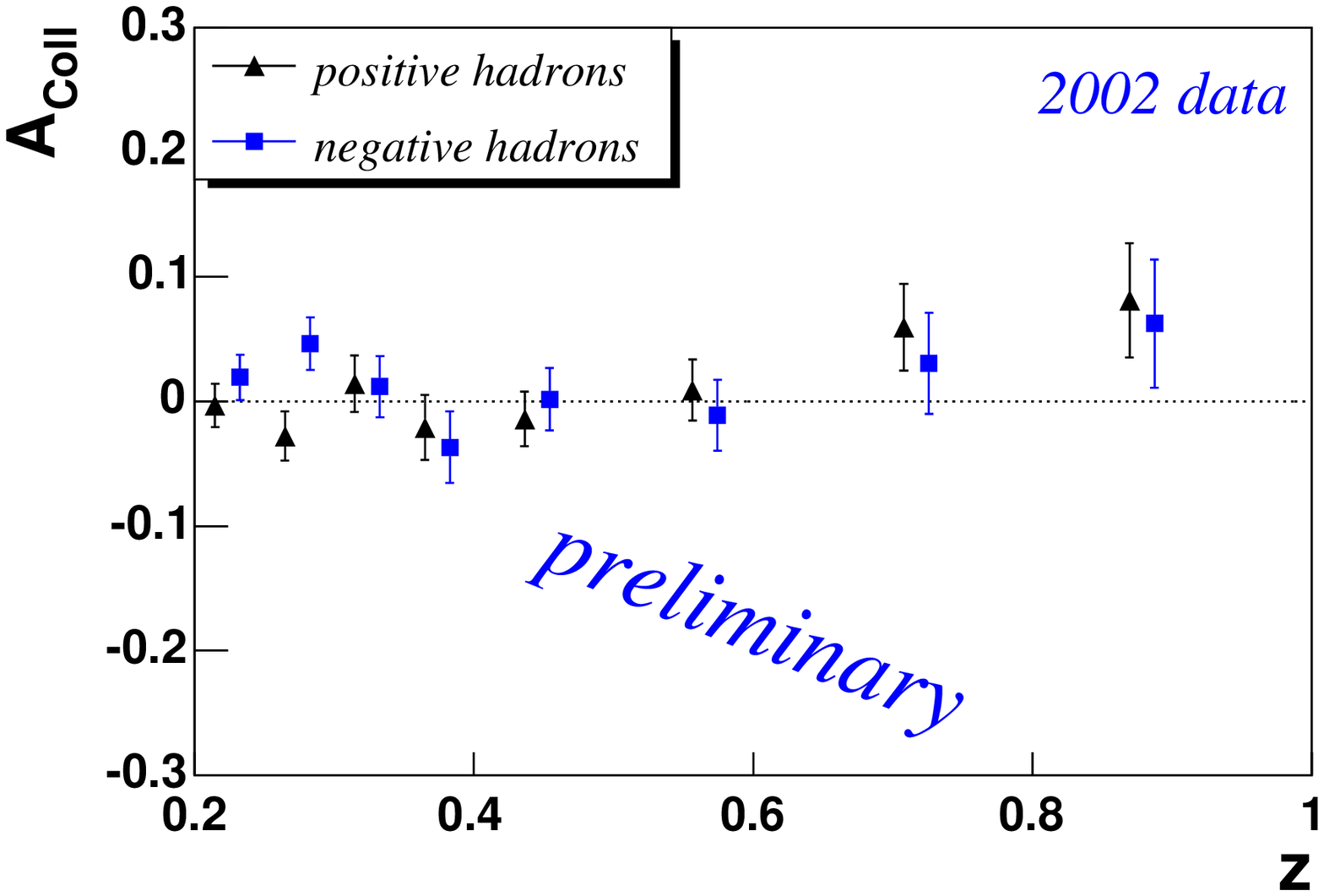}
}}
\hbox to 0pt{~}%
\vbox to \ht1{%
\centering{
\epsfxsize=\hsize\epsfbox{col_posneg_x_release_ah.eps}
\epsfxsize=\hsize\epsfbox{col_posneg_z_release_ah.eps}
}
}
\caption{%
     COMPASS 2002 deuteron data:\protect\cite{pagano} Collins asymmetries as function
     of $x_{Bj}$ (top) and of $z_h$ (bottom) for positive and negative hadrons.
     \label{fig:C_collins}}
\end{minipage}
\hfill
\begin{minipage}[t]{0.48\hsize}
%\setbox1=\vbox{%
%\centering{
%\epsfxsize=\hsize\epsfbox{col_posneg_x_release_ah.eps}
%\epsfxsize=\hsize\epsfbox{col_posneg_z_release_ah.eps}
%}
%}
\hbox to 0pt{~}%
%\vbox to \ht1{%
\centering{
\epsfxsize=\hsize\epsfbox{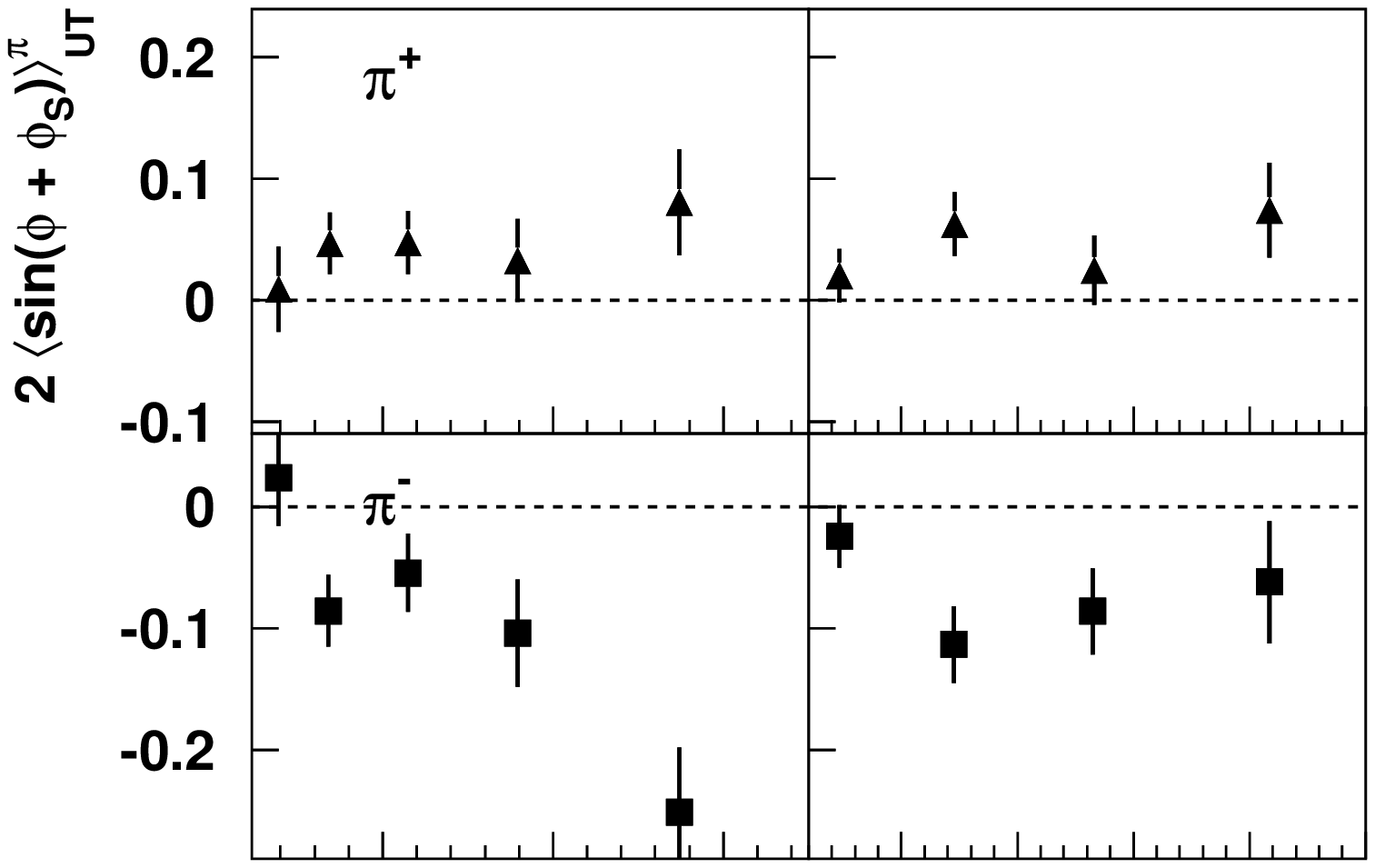}
\epsfxsize=\hsize\epsfbox{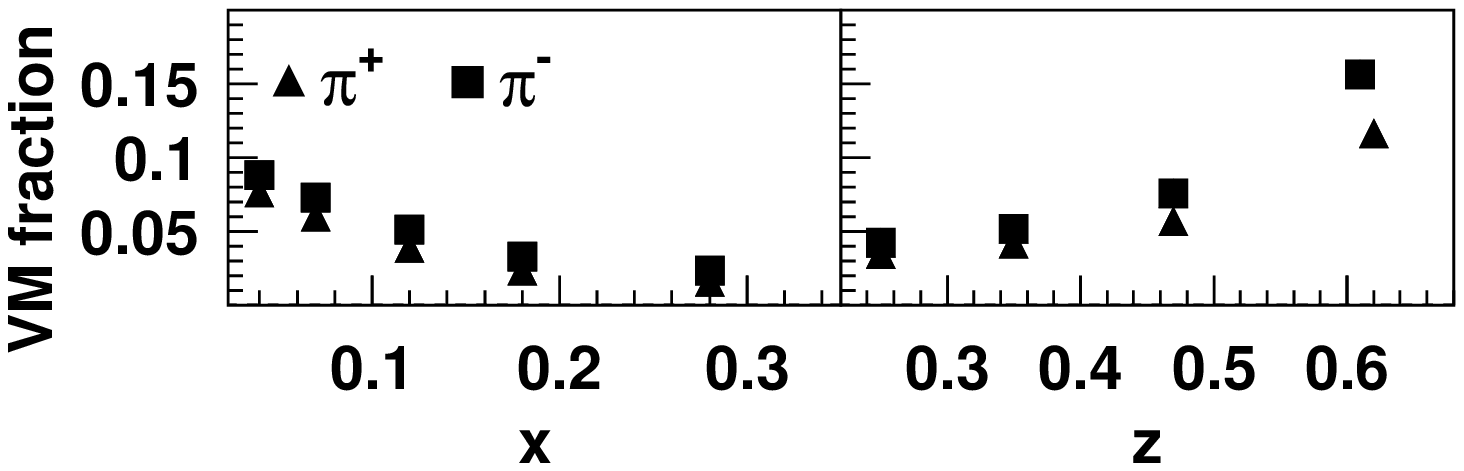}
}
%}
\caption{%
     HERMES 2002/3 proton data:\protect\cite{schnell} Collins asymmetries as 
     function of $x_{Bj}$ (left) and of $z_h$ (right) for positive (top) and negative (bottom) pions.
     The bottom panel shows the indicated the contribution from exclusively produced vector mesons from
     PHYTHIA Monte Carlo.
    \label{fig:H_collins}}
\end{minipage}
\end{figure}

Both, COMPASS\cite{pagano} and HERMES,\cite{schnell} presented first results on Collins and Sivers
asymmetries (Figs.~\ref{fig:C_collins} and \ref{fig:H_collins}). For the deuteron, both the Collins
and Sivers asymmetries, are compatible with zero while for the proton there is a hint of
positive values for favoured and of negative values for unfavoured fragmentation. For both experiments
more data are available and being analysed.
New quark polarimeters may help to get a better handle on the transversity distributions. The
experiments had already a first look\cite{dihadron} at the proposed di-hadron fragmentation,\cite{bacchetta} 
which avoids a $k_T$ convolution.
Drell--Yan processes in polarised pp collisions and even $\rm p \bar p$ collisions
at the future PAX experiment are other channels to look for transversity.\cite{efremov_ratcliffe}
 
New precise COMPASS data\cite{alexakhin} on the spin transfer to lambdas and anti-lambdas and the spin
density matrix in exclusive $\rho$ production are in good 
agreement with previous data. New data also were presented on transverse lambda polarisation.\cite{lambda_tr}
 
In the context of Generalised Parton Distributions
HERMES presented new results for the beam-charge asymmetry in deeply-virtual Compton-scattering off a deuteron 
target and for vector meson production.\cite{H_gpd}

\section{Outlook}
A wealth of new experimental results was presented at the Spin Symposium including first data from
COMPASS and RHIC. A precise measurement of the gluon polarisation in lepton--nucleon and in pp interactions 
is around the corner and first results were already presented.
%The continuous stream of data from HERMES is joined by new results from JLAB.
%A precise measurement of the gluon polarisation in lepton--nucleon and in pp interactions is around the 
%corner and first results were already presented.
The second missing piece in our understanding of the nucleon's spin structure is transversity and also 
here we saw first data from COMPASS and HERMES. The progress of the phenomenological models as well as 
that in NLO QCD theory will allow us to take full advantage of the wealth of experimental results.

\def\ibid{{\it ibid.}}
\def\etal{{\it et al.}}

\end{document}